# Report on Gamma-Ray Analysis of Seaweed Samples from Naturespirit Herbs LLC

Eric B. Norman[1,2], Keenan Thomas[1,2], Pedro Guillaumon[3], Alan R. Smith[2]

[1] Dept. of Nuclear Engineering, Univ. of California, Berkeley, CA

[2] Nuclear Science Division, Lawrence Berkeley National Laboratory, Berkeley, CA

[3] Dept. of Physics, Univ. of Sao Paulo, Sao Paulo, Brazil

Five samples of dried seaweed collected and prepared by Naturespirit Herbs LLC were delivered to UC Berkeley in January, 2014. Over a period of approximately 10 days, each of these samples was poured into a Marinelli beaker and individually gamma counted using a 46% relative efficiency high-purity germanium detector. The masses of the seaweed samples counted ranged from 299 to 349 grams. Counting times ranged from 23 hours to 68 hours for each sample. Gamma-ray spectra were acquired using the ORTEC Maestro system. The efficiency of the counting system was determined using a sample of Columbia River sediment prepared by the U. S. National Institute of Standards and Technology[1]. This material contains a calibrated amount of $^{137}$Cs ($t_{1/2}$ = 30.07 years), which is one of the fission products released as a result of the Fukushima reactor accident. The river sediment sample was also placed into a Marinelli beaker and counted on the same detector for a period of 49 hours. The decay of $^{137}$Cs produces a characteristic gamma ray at energy of 662 keV. Thus by comparing the counting rate of this gamma ray observed from a seaweed sample to that observed from the river sediment, the activity of $^{137}$Cs in the seaweed was determined. $^{134}$Cs ($t_{1/2}$ = 2.06 years) is another fission product that was released from Fukushima and emits characteristic gamma rays at energies of 604 and 795 keV. Searches were made for these gamma rays as well. In all five seaweed samples, no evidence of $^{134}$Cs was observed and thus only upper limits on the possible levels of this isotope were determined. Both $^{134}$Cs and $^{137}$Cs are produced in nuclear reactors and their ratio depends on how long the reactor was operating.

Approximately one month after the Fukushima reactor accident, a sample of weeds was collected in the Oakland hills. These weeds were immediately (April, 2011) counted and clear evidence of both $^{137}$Cs and $^{134}$Cs was observed.[2] These weeds were dried and preserved until the present. These weeds were counted again on the same detector used for the seaweed analysis. A portion of the gamma ray spectrum observed from these weeds (in February, 2014) is shown in Figure 1(a). The labelled peaks are those from the decays of $^{137}$Cs and $^{134}$Cs contained in the weeds. All of the other peaks are from naturally-occurring radioisotopes contained the building materials and shield surrounding the germanium detector. The combined activity of $^{137}$Cs and $^{134}$Cs contained in this 30.6-gram weed sample is 0.76 Bq. (Note that 1 Bq = 1 decay per second). Figure 1(b) shows the same energy region observed from the Naturespirit Kombu seaweed sample. As can be seen, a very small peak at 662-keV from $^{137}$Cs was seen, however, no evidence of $^{134}$Cs was observed. The fact that we did not observe $^{134}$Cs in the seaweed means that the radioactivity we observed did not come from Fukushima. The presence of $^{137}$Cs is probably due to atmospheric nuclear weapons tests done in the 1950's and 1960's.

The much more intense gamma ray observed (above background levels) from all the seaweed samples is that produced by the decay of the naturally occurring radioisotope $^{40}$K ($t_{1/2}$ = 1.27x10$^9$ years). In fact, from our measurements we determine that the potassium levels in these seaweeds vary from 1.5% to 12% by mass, and 0.0117% of all potassium is $^{40}$K. Thus observation of a low level of $^{40}$K activity in seaweed is to be expected. Our results for the activities of $^{137}$Cs and $^{40}$K in each of the seaweed samples are listed in Table 1.

The conclusion of this work is that no detectable level of radioactivity attributable to the Fukushima reactor accident was observed in any of the five seaweed samples. The observed levels of $^{137}$Cs in the seaweed are less than 1/4000$^{th}$ (i.e., less than 0.025%) of the Derived Intervention Level established by the U. S. Food and Drug Administration for $^{134+137}$Cs in food.[3]

**Acknowledgements**

This work was supported in part by the US Department of Energy National Nuclear Security Administration under Award No. DE-NA0000979.

**Table 1. Results of gamma counting Naturespirit Herbs LLC seaweed samples.**

| Sample Name | Collection Date | $^{137}$Cs Activity Bq/kg | $^{40}$K Activity Bq/kg |
|---|---|---|---|
| Kombu | July 25, 2013 | 0.244 ± 0.010 | 1318 ± 65 |
| Wakame | July 11, 2013 | 0.224 ± 0.012 | 1354 ± 65 |
| Kelp Fronds | June 12, 2013 | 0.219 ± 0.010 | 3816 ± 100 |
| Gigartina | July 26, 2013 | 0.020 ± 0.020 | 456 ± 20 |
| Bladderwrack | July 9, 2012 | 0.276 ± 0.010 | 1032 ± 30 |

Note: 1 Bq = 1 disintegration/second

FDA Derived Intervention Level

for $^{134}$Cs + $^{137}$Cs = 1200 Bq/kg in food

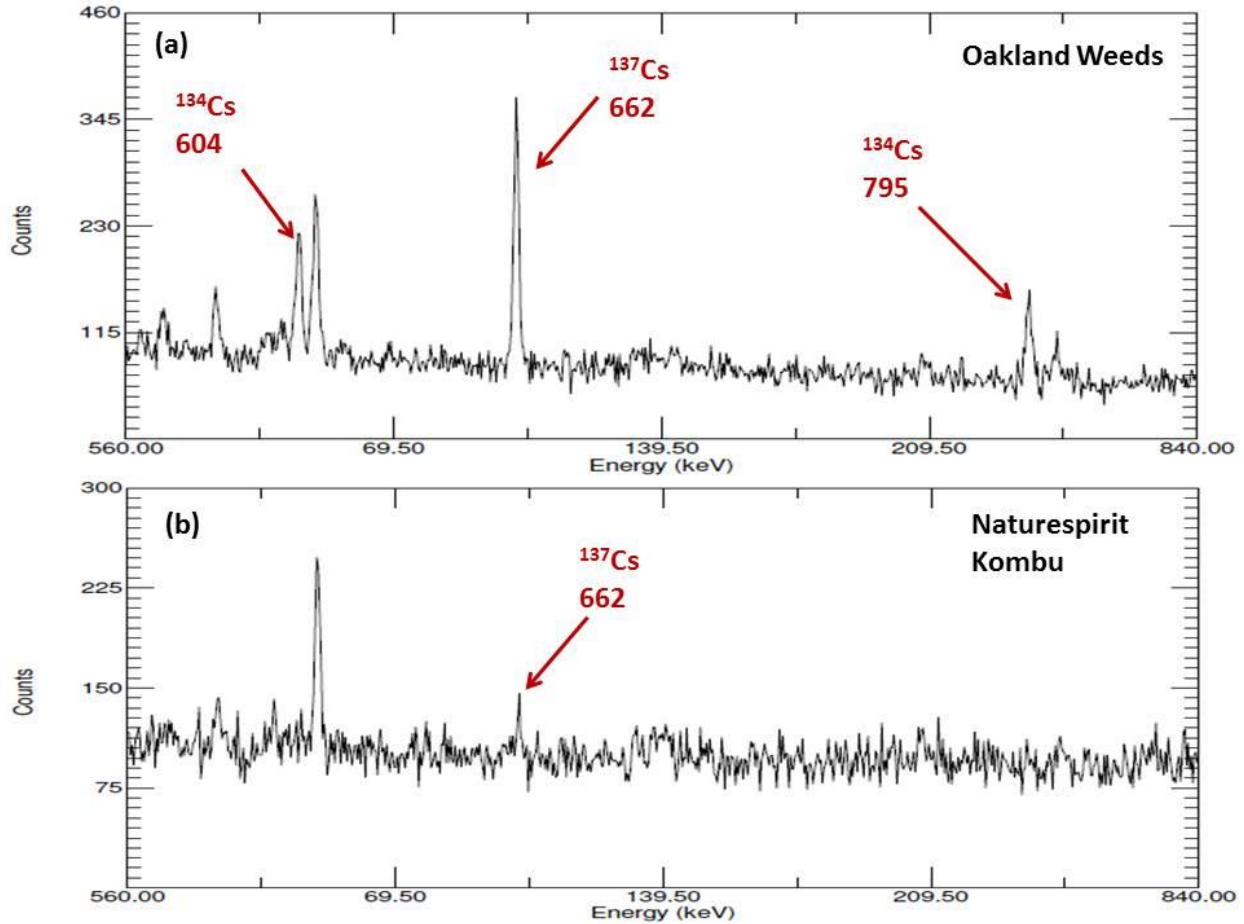

**Figure 1.** (a) Portion of the energy spectrum observed from counting 30.4 grams of dried weeds collected in the Oakland Hills in April 2011. Gamma-ray lines attributable to the decays of $^{134}$Cs and $^{137}$Cs are clearly seen. All other lines in the spectrum are from the decays of naturally occurring radionuclides. (b) Same portion of the energy spectrum observed from Naturespirit Herbs LLC's Kombu seaweed sample. No evidence of $^{134}$Cs is seen, but a small peak attributable to $^{137}$Cs is observed.